\documentclass[
journal=ancac3, 
manuscript=article]{achemso}

\usepackage[version=3]{mhchem} 
\usepackage[x11names]{xcolor}

 \author{Sen Shao}
\affiliation{Division of Physics and Applied Physics, School of Physical and Mathematical Sciences, Nanyang Technological University, 21 Nanyang Link, 637371, Singapore }

\author{Jia-Xin Yin}
\affiliation {Department of Physics, Southern University of Science and Technology, Shenzhen, Guangdong 518055, China}

\author{Ilya~Belopolski}
\affiliation {RIKEN Center for Emergent Matter Science (CEMS), Wako, Saitama 351-0198, Japan}

\author{Jing-Yang You}
\affiliation{Department of Physics, National University of Singapore, 2 Science Drive 3, Singapore 117551}

\author{Tao Hou}
\affiliation{Division of Physics and Applied Physics, School of Physical and Mathematical Sciences, Nanyang Technological University, 21 Nanyang Link, 637371, Singapore }

\author{Hongyu Chen}
\affiliation{Division of Physics and Applied Physics, School of Physical and Mathematical Sciences, Nanyang Technological University, 21 Nanyang Link, 637371, Singapore }

\author{Yuxiao Jiang}
\affiliation {Laboratory for Topological Quantum Matter and Advanced Spectroscopy (B7), Department of Physics, Princeton University, Princeton, New Jersey 08544, USA}

\author{Md Shafayat Hossain}
\affiliation {Laboratory for Topological Quantum Matter and Advanced Spectroscopy (B7), Department of Physics, Princeton University, Princeton, New Jersey 08544, USA}

\author{Mohammad Yahyavi}
\affiliation{Division of Physics and Applied Physics, School of Physical and Mathematical Sciences, Nanyang Technological University, 21 Nanyang Link, 637371, Singapore }

\author{ Chia-Hsiu Hsu}
\affiliation{Division of Physics and Applied Physics, School of Physical and Mathematical Sciences, Nanyang Technological University, 21 Nanyang Link, 637371, Singapore }

\author{Yuan Ping Feng}
\affiliation{Department of Physics, National University of Singapore, 2 Science Drive 3, Singapore 117551}

\author{Arun Bansil}
\affiliation{Department of Physics, Northeastern University, Boston, MA 02115, USA}

\author{M.~Zahid~Hasan}
\affiliation {Laboratory for Topological Quantum Matter and Advanced Spectroscopy (B7), Department of Physics, Princeton University, Princeton, New Jersey 08544, USA}

\author{Guoqing Chang}
\email{guoqing.chang@ntu.edu.sg}
\affiliation{Division of Physics and Applied Physics, School of Physical and Mathematical Sciences, Nanyang Technological University, 21 Nanyang Link, 637371, Singapore }

\title[\texttt{achemso} demonstration]
{Intertwining of magnetism and charge ordering \\ in kagome FeGe}

\begin{document}

\newpage

\begin{abstract}
Recent experiments report a charge density wave (CDW) in the antiferromagnet FeGe, but the nature of the charge ordering and the associated structural distortion remains elusive. We discuss the structural and electronic properties of FeGe. Our proposed ground state phase accurately captures atomic topographies acquired by scanning tunneling microscopy.  We show that the 2$\times$2$\times$1 CDW likely results from the Fermi surface nesting of hexagonal-prism-shaped kagome states. FeGe is found to exhibit distortions in the positions of the Ge atoms instead of the Fe atoms in the kagome layers. Using in-depth first-principles calculations and analytical modeling, we demonstrate that this unconventional distortion is driven by the intertwining of magnetic exchange coupling and CDW interactions in this kagome material. Movement of Ge atoms from their pristine positions also enhances the magnetic moment of the Fe kagome layers. Our study indicates that magnetic kagome lattices provide a material candidate for exploring the effects of strong electronic correlations on the ground state and their implications for transport, magnetic, and optical responses in materials.
\end{abstract}


\section{Introduction}

Density wave orders in two-dimensional and layered materials are known to exhibit rich interplay with their underlying lattice motifs\cite{Mudry_Kekule, Pasupathy_Kekule,ShuyunZhou_Kekule,Hasan_Wely,Yin_topo,HongJunGao_CsV3Sb5, Zeljkovic_CsV3Sb5, XianhuiChen_CsV3Sb5,Jiang_Uncon,Li_Observa,Tan_Charg,Neupert_Charg,Yu_Unusual}. For example, the honeycomb lattice exhibits the kekul\'e bond-density-wave order\cite{Mudry_Kekule, Pasupathy_Kekule}. In graphene, such kekul\'e order can give rise to massive Dirac fermions via chiral symmetry breaking, a mechanism analogous to dynamical mass generation of fundamental fermions in the Standard Model \cite{ShuyunZhou_Kekule}. Similarly, the kagome lattice supports a rich tapestry of charge density wave (CDW) orders \cite{HongJunGao_CsV3Sb5, Zeljkovic_CsV3Sb5, XianhuiChen_CsV3Sb5,Jiang_Uncon,Li_Observa,Tan_Charg,Neupert_Charg,Yu_Unusual}. In KV$_3$Sb$_5$, the CDW is associated with time-reversal-breaking orbital current order, closely related to the Haldane model for Chern insulators and the Varma model for the cuprate superconductors \cite{Zurab_KV3Sb5}. The exploration of new quantum materials exhibiting density wave orders under broken time-reversal symmetry, especially in the honeycomb and kagome lattices, promises the discovery of exotic many-body quantum phases, including the exotic orbital current orders\cite{Yang_Inter}, unconventional topological superconductors\cite{Zurab_KV3Sb5,Mielke_Nodel,Jiang_Unconv} and phases supporting fractional excitations \cite{Mudry_Kekule}.

The antiferromagnet FeGe assumes a crystal structure comprised of alternating honeycomb and kagome atomic layers. Recent neutron scattering, spectroscopy and transport measurements suggest a CDW in FeGe, providing an example of a CDW in a kagome magnet \cite{PengchengDai_FeGe, Jiaxin_FeGe}. However, the nature of the structural distortion and the mechanism driving the CDW remain elusive. Here, by analyzing in-depth first-principles calculations with optimized atomic structure, we predict that FeGe hosts an unusual hybrid kagome-honeycomb CDW. We find a $2\times2\times1$ CDW, which takes the form of a generalized kekul\'e distortion of the Ge honeycomb layers, while the Fe kagome layers almost remain unchanged. Our analysis indicates that the honeycomb generalized kekul\'e distortion is likely to be driven by the Fermi surface nesting of kagome electronic states. Our predicted honeycomb-kagome CDW is further substantiated by the matching with the atomic distribution of the electronic density of states data obtained via scanning tunneling microscopy (STM).

\subsection{Results and discussion}
\begin{figure}[!hbtp]
	\centering
	\includegraphics[width=0.7\textwidth]{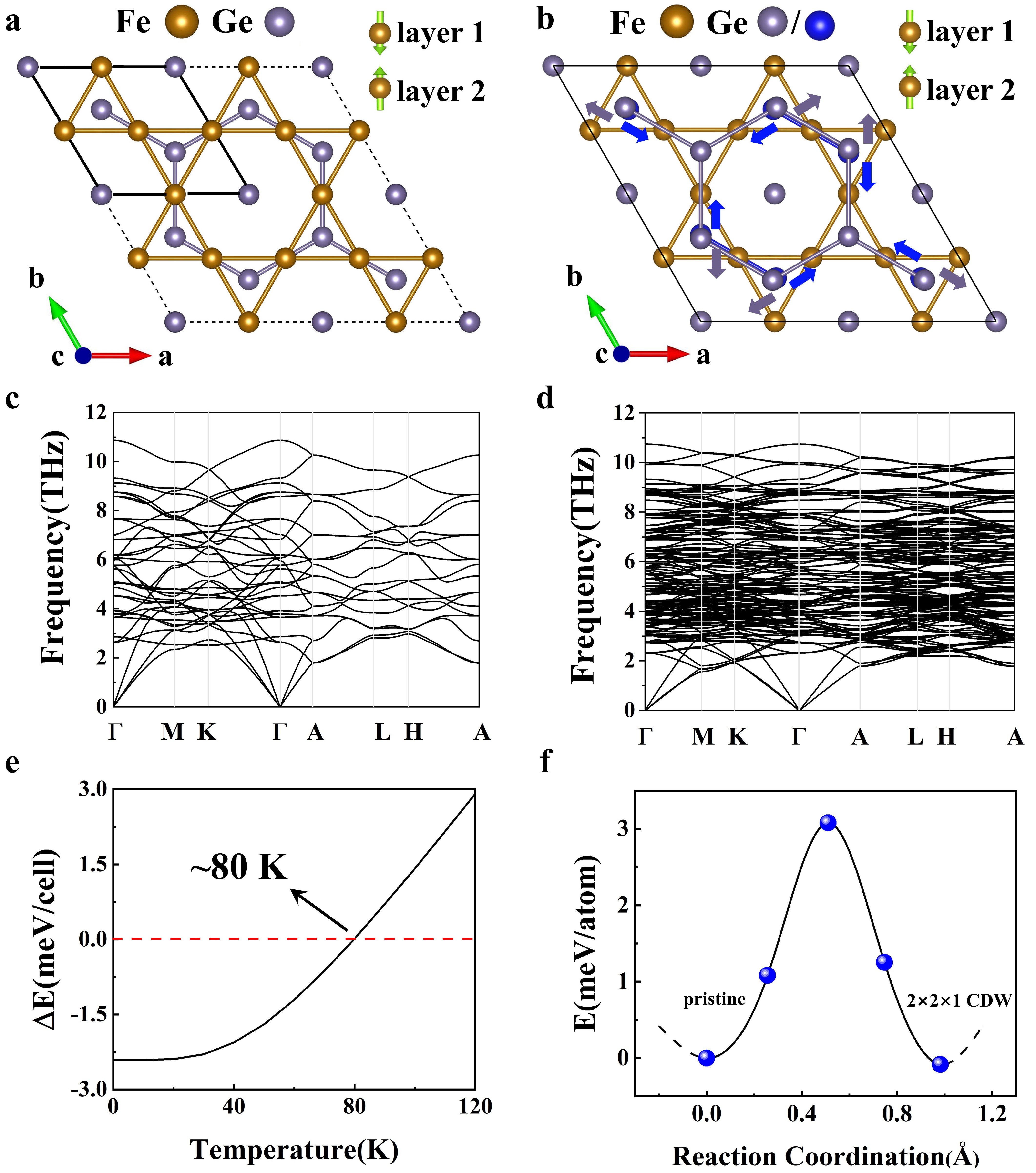}\\
	\caption{(a) A 2$\times$2$\times$1 supercell of the pristine phase, where the primitive cell is marked with the black solid line. (b) The 2$\times$2$\times$1 CDW phase, where the blue and purple arrows show the in-plane movement of Ge atoms on the top and middle Ge layers, respectively. The out-of-plane movement of the Ge atoms is given in Note 1 (Supporting Information). The green arrows in (a) and (b) represent the direction of the magnetic moment in different kagome layers. (c), (d) Phonon spectra of pristine and 2$\times$2$\times$1 CDW phases. (e) Gibbs free energy of 2$\times$2$\times$1 CDW phases per cell (48 atoms) relative to the pristine phase. (f) Reaction path between the pristine and 2$\times$2$\times$1 CDW phase calculated by NEB\cite{Henkel_neb1,Mills_neb,Shepp_neb}. } \label{fig1}
\end{figure}

The pristine phase of FeGe is in the space group $\textit{P}6/\textit{mmm}$ (No.191) [Figure ~\ref{fig1}a], where Fe atoms form kagome layers. There are two types of Ge atoms, one is located at the hexagonal center of the Fe kagome layer and the other forms honeycomb layers located in between the two kagome layers.  Our calculation shows that the pristine phase is inter-layer antiferromagnetic with magnetic moment perpendicular to the kagome layer. By using the high-performance structure research software CALYPSO\cite{Wang_calypso1,Wang_calypso2,Gao_calypso}, we identify four types of antiferromagnetic FeGe supercell structures (Note 1 in Supporting Information). Among these, the 2$\times$2$\times$1 and 2$\times$2$\times$2 supercell structures are the lowest energy structures that compete with the pristine phase after the zero-point energy is included. This is consistent with experimental results\cite{PengchengDai_FeGe, Jiaxin_FeGe}, where the 2$\times$2$\times$1 and 2$\times$2$\times$2 CDW interactions have been reported. Our ground state 2$\times$2$\times$1 structures are shown in Figure~\ref{fig1}b. The 2$\times$2$\times$1 structure shares the $\textit{P}6/\textit{mmm}$ space group with the pristine phase. Arrows indicate the movement of atoms in the supercell in the ab-plane. The Fe kagome layers are seen to be almost unchanged. The Ge atoms, which form the honeycomb lattice in different layers, move in opposite directions. As a result of this distortion, we have two inequivalent Fe atoms and six inequivalent Ge atoms in the 2$\times$2$\times$1 supercell.

To understand the transition between the phases, we plot the computed phonon band structure of the pristine and the 2$\times$2$\times$1 CDW phase in Figure~\ref{fig1}c,d, respectively. Both these phases are seen to be dynamically stable. This should be contrasted sharply with most of the previously reported materials, which display a CDW phase transition, where the phonon spectrum of the pristine phase exhibits a softening mode at low temperature\cite{Weber_Extend,Diego_van,Kvashnin_coexi,Xi_Strongly}. To further understand the nature of the phase transition, we consider the Gibbs free energy of 2$\times$2$\times$1 CDW phase relative to the pristine phase, which is expected to increase with increasing temperature (Figure~\ref{fig1}e). At low temperatures, the energy difference is negative, but it changes sign at 80 K, where the phase transition takes place. This is close to the results of the latest experiments\cite{PengchengDai_FeGe,Jiaxin_FeGe}, where the CDW transition is reported at 100-110 K. We further consider the energy barrier in the reaction pathway between the pristine and the 2$\times$2$\times$1 CDW phase in Figure~\ref{fig1}f, where both phases are local minima in energy and the CDW phase shows slightly lower energy. There is an energy barrier of $\sim$3.06 meV/atom along the reaction path. This is different from the case of the kagome family AV$_3$Sb$_5$ (A = K, Rb, Cs) \cite{Tan_Charg}, where the pristine phase sits at a saddle point of a potential energy surface, so that the softening mode will trigger a spontaneous change to a lower energy phase.

\begin{figure}[!tb]
	\centering
	\includegraphics[scale=0.6,angle=0]{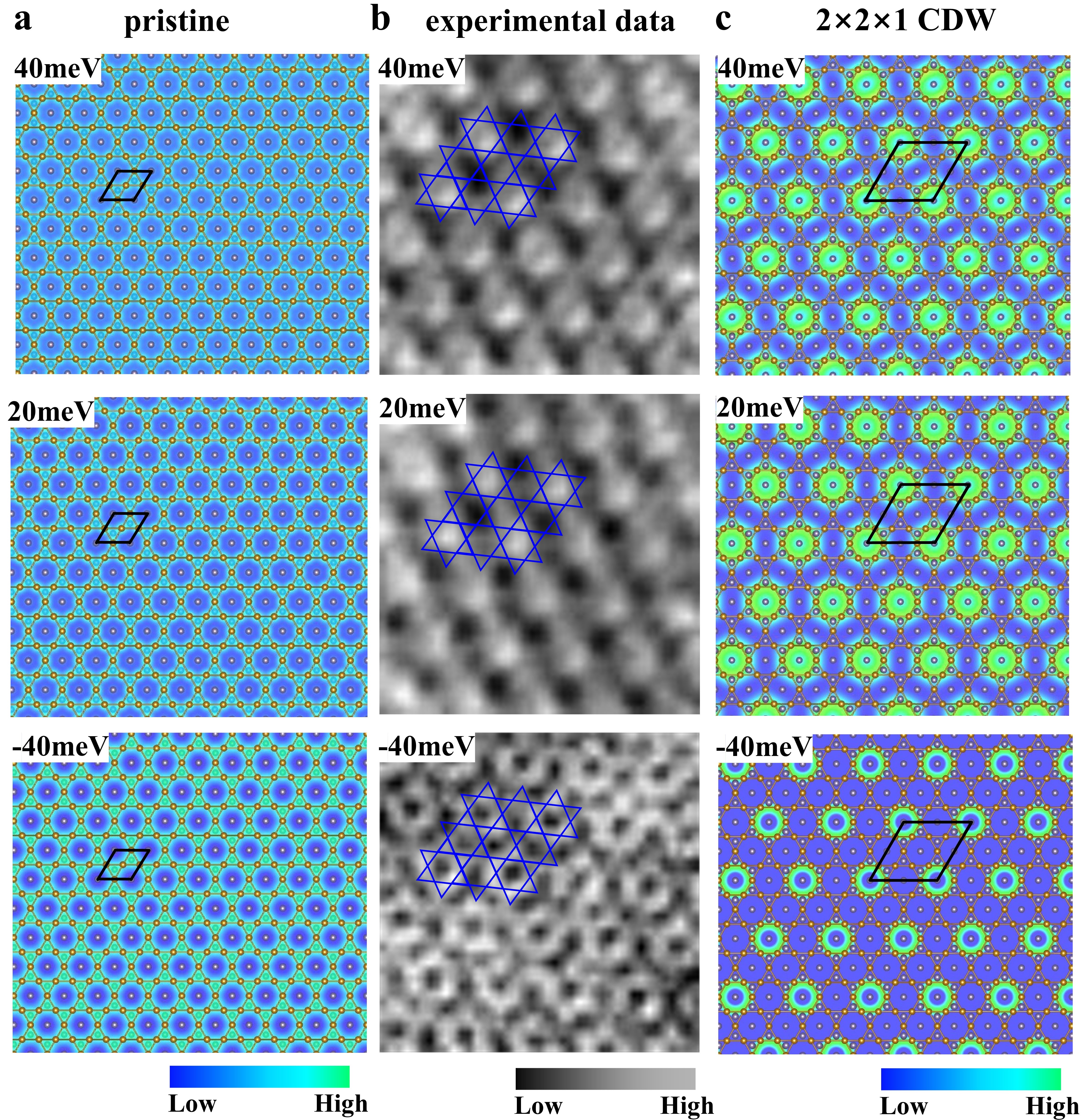}\\
	\caption{(a) DFT simulation of the surface charge distribution on the kagome layer of the pristine phase, where the black parallelogram marks the primitive cell. (b) Tunneling current maps measured by STM, where the blue lines mark the kagome lattice. (c) Simulated surface charge distribution on the kagome layer of the 2$\times$2$\times$1 CDW phase, where the black parallelogram marks the primitive cell. Upper, middle, and lower panels in (a)-(c) are the experimental STM images or the simulated surface charge distribution at 40, 20, and -40 meV relative to the Fermi energy, respectively.}\label{fig2}
\end{figure}

To further prove that the distorted 2$\times$2$\times$1 CDW phase correctly captures the experimentally observed CDW ground state, we show our STM measurements of the density of states and compare these with our theoretical predictions. Figure ~\ref{fig2}a simulates the charge distribution of the pristine phase at the Fe kagome surface.  The charge is mainly distributed on the Fe-triangles and its distribution will not change significantly as the energy moves around the Fermi level of the pristine phase. In contrast, in the 2$\times$2 CDW phase, STM measurements show a clear Fermi-level-dependent charge distribution (Figure ~\ref{fig2}b). At 20 and 40meV above the Fermi energy, the charge is mainly distributed in the Fe-hexagon and its adjacent Fe-triangles, but not in its neighboring Fe-hexagons.  At 40 meV below the Fermi energy, the center of the bright Fe-hexagons becomes dimmer, and the charge is mainly distributed around the boundary of the Fe-hexagons and Fe-triangles. The surface charge distribution based on the 2$\times$2$\times$1 CDW phase presented in Figure ~\ref{fig2}c captures the behavior seen in STM plots. These results show that our predicted 2$\times$2$\times$1 CDW phase correctly captures the distorted 2$\times$2 charge ordering structure observed experimentally in FeGe. 

\begin{figure}[!b]
	\centering
	\includegraphics[width=0.6\textwidth]{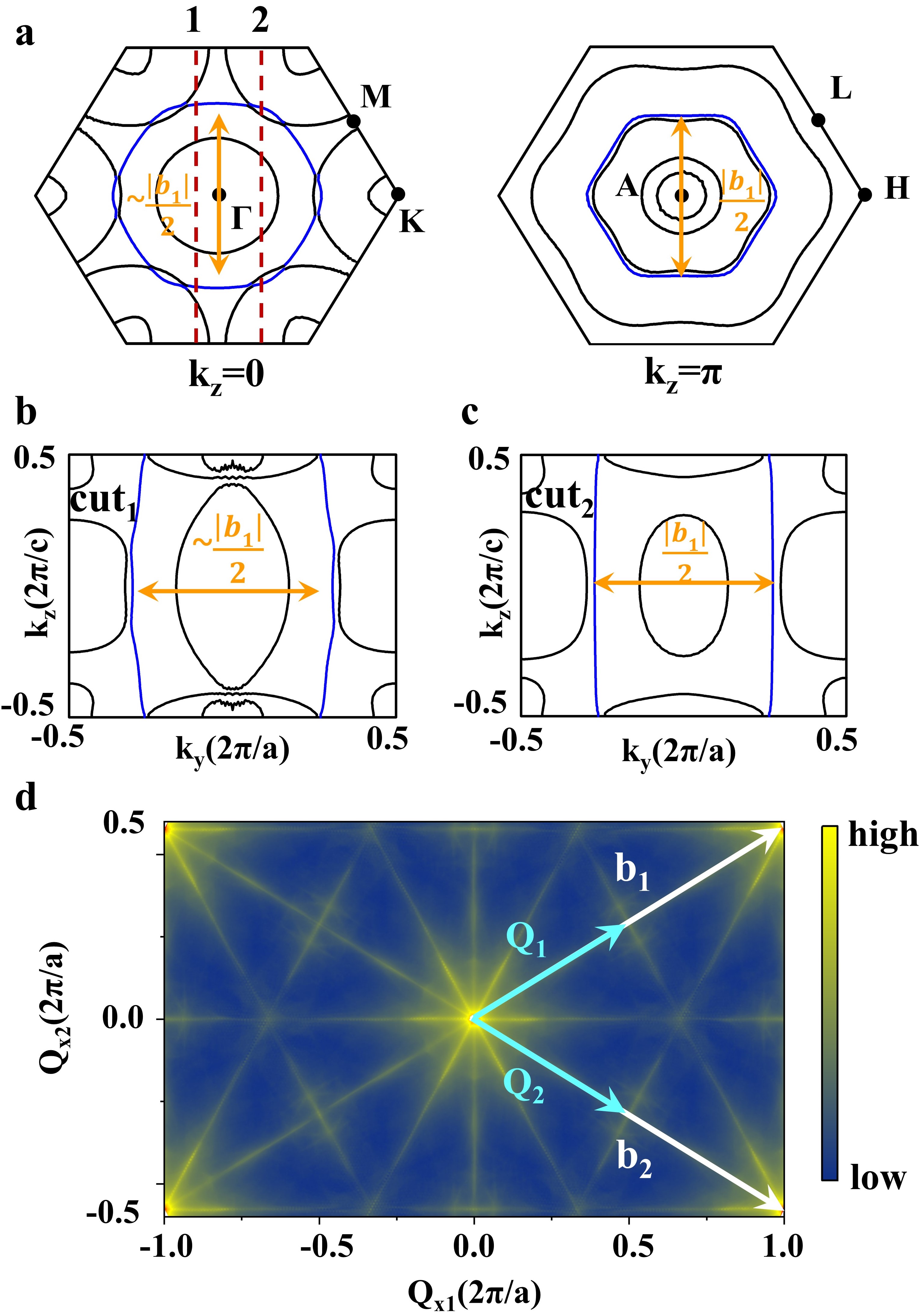}\\
	\caption{(a) Two-dimensional Fermi surface profiles of pristine phase in the ${k_{z}}$=0 and ${k_{z}}$=$\pi$ planes. (b) Fermi surface contour in the $yz$ plane along the cut$_{1}$ (${k_{x}}$=-${\sqrt{3}\pi}$/12) in panel (a). (c) Fermi surface contour in the $yz$ plane along the cut$_{2}$ (${k_{x}}$=${\sqrt{3}\pi}$/6) in panel (a). (d) Normalized Lindhard response function for the pristine phase in the $\textit{b}$$_{1}$-$\textit{b}$$_{2}$ plane.}\label{fig3}
\end{figure}

We turn now to discuss the origin of the CDW interaction that induces atomic distortions. In this connection, we first checked the Fermi surface of the pristine phase (Figure~\ref{fig3}a-c). Interestingly we observed a very similar hexagonal Fermi surface both at  ${k_{z}}$=0 and ${k_{z}}$=$\pi$ planes (colored blue in Figure~\ref{fig3}a). This indicates that the colored states have very small dispersion along the $z$-direction and that a hexagonal prismatic shape is present in the Fermi surface (Figure S5). The Fermi surface in the $yz$ plane also shows this behavior (Figure~\ref{fig3}b,c). The momentum separation of the quasi-one-dimensional parallel sides of the hexagonal-prism-shaped Fermi surface is about half of the unit vector of the pristine phase, as indicated by the yellow arrows in Figure~\ref{fig3}a-c. To further demonstrate the role of Fermi-surface nesting in the CDW formation, we calculate the zero frequency susceptibility ${\chi^{0}(q,\omega = 0)}$, i.e. the Lindhard response function\cite{Lindard_on,Borisenko_Pse,Borisenko_Two,Dugdale_Life}, to determine the nesting vector. The two-dimensional renormalized Lindhard function is plotted in Figure~\ref{fig3}d. Indeed the maxima are located at the center of the boundary of Brillouin zone (BZ) with the nesting vectors ${\textbf{\textit{Q}}_{1}}$ = (0.5, 0) and $\textbf{\textit{Q}}_{2}$=(0, 0.5). This is in accord with the corresponding experimental observations\cite{PengchengDai_FeGe,Jiaxin_FeGe}. These results indicate that the Fermi-surface nesting is the likely driver of the 2$\times$2$\times$1 CDW phase. The role of electron-phonon coupling in this connection, however, deserves further attention.

\begin{figure*}[!htb]
	
	\begin{center}
		\includegraphics[width=1.0\textwidth]{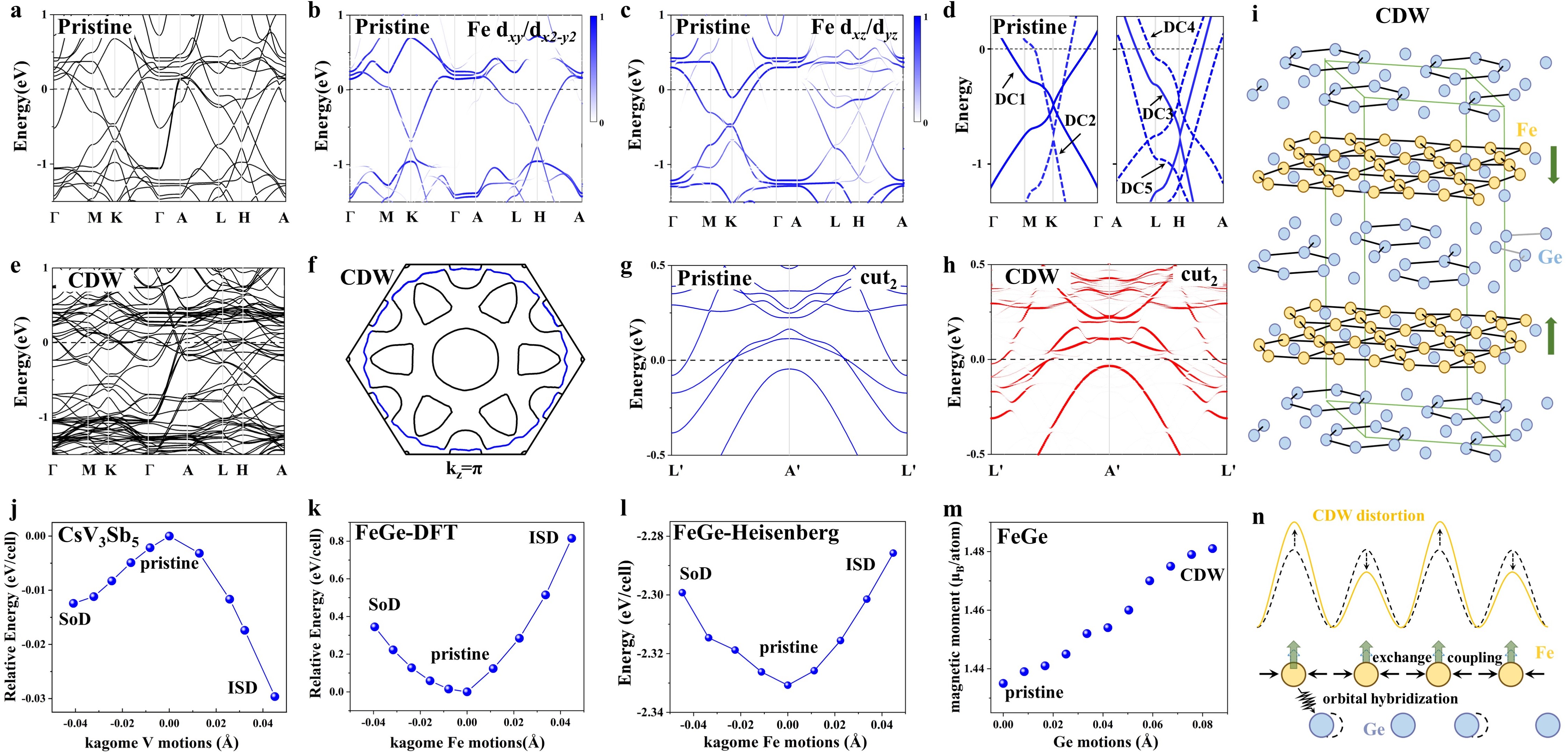}
	\end{center}
	\caption{(a) Band structure of the pristine phase including SOC effects. (b), (c) Orbital-resolved band structure of $\textit{d$_{xy}$}$/$\textit{d$_{x^2-y^2}$}$ and $\textit{d$_{xz}$}/$\textit{d$_{yz}$} for the pristine phase, respectively. (d) Schematic diagram of Dirac cones in the pristine phase, where the solid lines mark the bands that play a major role in Fermi surface nesting. (e) Band structures of the 2$\times$2$\times$1 CDW phase including SOC effects. (f) Constant energy contours for the 2$\times$2$\times$1 CDW phase in the k$_{z}$=$\pi$ plane at 10 meV below the Fermi energy. (g) Band structures of the pristine phase and (h) the unfolded band structures of the 2$\times$2$\times$1 CDW phase along the cut$_{2}$ plane along L$^{\prime}$-A$^{\prime}$-L$^{\prime}$, where L$^{\prime}$ is (0.166, 0.4167, 0.5) and A$^{\prime}$ is (0.166, -0.0833, 0.5). The CDW gap is located at (0.166, 0.1667, 0.5) with an energy gap of $\sim$45 meV (i) A schematic of the CDW distorted kagome antiferromagnet. The shortest bonds in the top Ge layer form the O-shaped bond textures, and the shortest bonds in the middle Ge layer form the Y-shaped bond textures. Orange and blue spheres represent Fe and Ge atoms, respectively. Dark green arrows represent magnetic directions. (j) Relative energy of CsV$_{3}$Sb$_{5}$ as a function of the average V motion. (k) Relative energy of FeGe as a function of the average Fe motion. (l) Hamiltonian of the Heisenberg model as a function of the average Fe motion. (m) Average magnetic moment of Fe as a function of the average Ge motion. (n) Sketch of the intertwining between magnetic exchange coupling and CDW interaction. The black dashed curve gives the charge density in the pristine phase. The orange curve gives the charge density in the CDW phase. Black arrows represent the magnetic exchange coupling effect on the Fe atoms. Green and white arrows represent the kagome-Fe magnetic moment in the CDW phase and the pristine phase, respectively.}
	\label{fig4}
	
\end{figure*}

To gain further insight into the Fermi surface nesting scenario, we consider the electronic band structure of the pristine phase. Figure~\ref{fig4}a depicts the electronic band structure of the pristine phase with SOC. The Fe-$d$ orbital-projected band structures to focus on the singularity of the kagome lattice are shown in Figure~\ref{fig4}b,c. The 'flat bands' are seen to lie slightly above the Fermi energy with fairly large bandwidths, which is consistent with previous studies\cite{Setty_Electron}. We identify five Dirac cones (DCs) derived from the kagome lattice, which display van Hove singularities at the M/L points and Dirac points at the K/H points. A schematic diagram of these Dirac cones is shown in Figure~\ref{fig4}d. Notably, DC2 and DC3 possess similar dispersions, but the related van Hove singularities lie at different energies. This is because the $\textit{d$_{xy}$}/$$\textit{d$_{x^2-y^2}$}$ orbitals, which dominate DC2 and DC3, lie in the kagome plane, and their hybridization with the upper and lower Ge atoms is weak. In contrast, the other three Dirac cones are mainly contributed by the out-of-plane $\textit{d$_{xz}$}/$\textit{d$_{yz}$} orbitals, and due to their strong hybridization with Ge orbitals, they show different dispersions in the ${k_{z}}$=0 and ${k_{z}}$=$\pi$ planes. The DC1 and DC3 Dirac cones shown by the solid-blue lines in Figure \ref{fig4}d play a major role in producing the hexagonal prism highlighted in Figure~\ref{fig3}. The kagome-derived DC1 plays a major role in the hexagonal Fermi surface in the ${k_{z}}$=0 plane, while the main contribution to the hexagonal Fermi surface gradually becomes DC3 as the 2D Fermi surface lifts from ${k_{z}}$=0 to $\pi$.   

Now we discuss the effects of CDW interaction on the band structure. The E-k dispersion in the 2$\times$2$\times$1 CDW phase with the inclusion of SOC is given in Figure ~\ref{fig4}e. Most kagome features: the Dirac points and the flat bands can still be observed in the 2$\times$2$\times$1 CDW phase. The main effects of the CDW interactions are on states that are nested by the CDW Q-vectors. Taking the constant energy contour at k$_{z}$=$\pi$ at 10 meV below the Fermi energy as an example, the blue hexagonal contour in the pristine phase changes into a circle-like contour in the 2$\times$2$\times$1 CDW phase (Figure \ref{fig4}f and S7). This evolution of the constant energy contours weakens Fermi surface nesting and induces CDW instability. Energy dispersion along with L$^{\prime}$-A$^{\prime}$-L$^{\prime}$ in cut$_{2}$ plane of the pristine phase is shown in Figure \ref{fig4}g, and the unfolded band structure of the 2$\times$2$\times$1 CDW phase along the same path is shown in Figure \ref{fig4}h. A clear gap-opening due to CDW interactions can be seen along this path.

We emphasize that the Fermi surface nesting comes from the kagome-derived Dirac cones, but structural distortions originate from motions of the Ge atoms, which lead to the formation of Ge bonds along the c axis and generalized kekul\'e distortions\cite{kekule_Stu,Chamon_Soliton,Herrera_Elec} in the honeycomb layer: the resulting shortest bonds form an "O"/"Y" shaped pattern as shown in Figure \ref{fig4}i.  In the nonmagnetic kagome material AV$_{3}$Sb$_{5}$ (A=K, Rb, and Cs), the star of David (SoD) and the inverse star of David (ISD) distortion of the kagome layer decrease the total energy (Figure ~\ref{fig4}j). However, in FeGe, the Fermi surface nesting related to Fe atoms drives the motion of the Ge atoms instead. This unconventional distortion is caused by the intertwining of magnetism and CDW interactions.  Our analysis shows that the SoD/ISD distortion in the Fe kagome layers will increase the total energy (Figure \ref{fig4}k). Our Heisenberg model-based calculations (Note 3 in Supporting Information) yield a similar behavior (Figure ~\ref{fig4}l). These results show that the magnetic exchange coupling tends to force the Fe atoms to remain in their original positions in the kagome layer, while the CDW interaction tends to move the Fe atoms away from their original positions. As a consequence of these two competing tendencies, the Fermi surface nesting forces are transmitted to the Ge atoms through the hybridization of  Fe and Ge orbitals (Figure S10), resulting in the motion of Ge atoms in the 2$\times$2$\times$1 CDW phase. 

In the latest experiments, the observed enhancement of the magnetic moment may be explained by the chiral flux phase of circulating currents and the effects of electronic correlations independent of Fermi surface nesting\cite{PengchengDai_FeGe}. Our calculations show that the magnetic moment of the Fe atom in the 2$\times$2$\times$1 CDW phase is 1.48 $\mu_B$, which is greater than the value of 1.44 $\mu_B$ in the pristine phase. This magnetic moment enhancement found in our work is consistent with aforementioned experimental observations. To better understand this effect, we calculated the spin-resolved partial density of states associated with Fe atoms. In the 2$\times$2$\times$1 CDW phase, the spin-up (spin-down) states increase (decrease) in the top kagome layers and behave in an opposite manner in the bottom Fe-kagome layer, which leads to the magnetic enhancement of Fe atoms (Figure S11). To analyze whether such an enhancement is related to Ge motions, we calculated the change of magnetic moment with the size of Ge motions. As shown in Figure 4m, the magnetic moment increases approximately linearly with Ge displacement indicating that the motions of Ge atoms can lead to greater spin splitting of Fe atoms, which, in turn, enhances the magnetic moments.

\section{Conclusion}

In summary, we have investigated the structural and electronic properties of antiferromagnetic kagome metal FeGe. Our in-depth analysis of the ground state 2$\times$2$\times$1 CDW phase indicates that this phase will transform into the pristine phase at a temperature of $\sim$80 K, which is comparable to the corresponding experimentally measured transition temperature of $\sim$100-110 K\cite{PengchengDai_FeGe,Jiaxin_FeGe}. Our predicted 2$\times$2$\times$1 CDW phase accurately captures the salient topographic and spectroscopic features our STM/STS experiments and the experimentally observed enhancement of the magnetic moment\cite{PengchengDai_FeGe,Jiaxin_FeGe}. Our study unveils the presence of an unconventional intertwining between magnetism and CDW interactions in kagome magnets (Figure \ref{fig4}n), which forces Fe atoms of the kagome layers to remain at the original positions, but leads to the motion of Ge atoms and the formation of a generalized kekul\'e distortion in the Ge honeycomb layers. Since Kekul\'e distortions in graphene open conduction channels between the valleys that influence optical absorption and conductivity\cite{Herrera_Elec}, we expect the generalized kekul\'e distortions in FeGe to also yield interesting physical effects. Our study provides insight into the CDW phase and indicates that further exploration of the topological nature of the ground state of magnetic kagome-honeycomb lattices and their transport and optical responses will be interesting. We also identify the structure of the 2$\times$2$\times$2 CDW  phase and the structures of the two related metastable 1$\times$${\sqrt{3}}$$\times$1 and 2$\times$2${\sqrt{3}}$$\times$1 phases (Table S1), which could provide theoretical support for future experiments.






\section{Methods}
The CDW phases are proposed based on the structure search method implemented in the CALYPSO \cite{Wang_calypso1,Wang_calypso2,Gao_calypso} code. First, we fix the positions of the Ge atoms in the volume of the 2$\times$2$\times$1 pristine phase and generate positions of the Fe atoms randomly. Then, we fix the positions of the Fe atoms and generate the positions of Ge atoms randomly. After generating such 2$\times$2$\times$1 structures, we relax them using first-principles calculations using the Vienna ab-initio Simulation Package (VASP)\cite{VASP_ref} within the Perdew-Burke-Ernzerhof exchange-correlation functional\cite{PBE_ref}. The projector augmented-wave method \cite{PAW_ref} is implemented with 3$\textit{d}^{7}$4$\textit{s}^{1}$ and 4$\textit{s}^{2}$4$\textit{p}^{2}$ treated as valence electrons for Fe and Ge atoms, respectively. The cut-off energy of 520 eV and a dense k-point sampling of the Brillouin zone with a spacing of 2$\pi$ $\times$ 0.03 \AA$^{-1}$ are used to ensure the convergence of enthalpy within 1 meV/atom. The phonon spectra are calculated with the frozen phonon method and the Gibbs free energy is simulated in the framework of quasi-harmonic approximation as employed in PHONOPY \cite{phonopy_ref} code. The post-processing of band and DOS calculations is performed by the VASPKIT tool \cite{vaspkit_ref}.
The pristine phase of FeGe is found to be dynamically stable at 0 K, indicating that the CDW transition is induced by electrons nested at the Fermi level. Therefore, we focus on the electronic contribution to the CDW instability and interpret our results using a simple picture of the Fermi-surface nesting vectors. We also carry out an accurate calculation of the electronic susceptibility with $\omega$=0 Lindhard response function. In the constant-matrix-element approximation,\cite{Lindhard_ref1,Dugdale_Life} $\chi$$^{0}$(q,0) is given by:
          \begin{center}
          	 $\chi$$^{0}$(q,0) = $\sum_{K}$$\frac{f_{K}(1-f_{K+q}}{\epsilon(K+q)-\epsilon(K)}$
          \end{center}

Here, the sum is over the whole first Brillouin zone. The Fermi-Dirac distribution is: f$_K$=$\frac{1}{\frac{\epsilon(K)}{T}+1}$
The STM images of the 2$\times$2$\times$1 CDW and the pristine phases are simulated using a slab model, which has been used previously to simulate STM images in AV$_{3}$Sb$_{5}$ (A=K, Rb, Cs) \cite{Tan_Charg}. The STM images for the Fe kagome surface of pristine and 2$\times$2$\times$1 CDW phases are simulated using a 3-unit-cell thick slab (144 atoms) with a vacuum layer of $\sim$10 \AA. The STM images are calculated with a fixed height of $\sim$3 \AA $ $ above the surface in the energy range of ± 25 meV relative to the given energy. Structural relaxation and SOC effect are not considered in the STM simulations.

\section{Author Contributions}
S. S., J.-X. Y., I. B., and J.-Y. Y. contributed equally to this work. G.C. was responsible for the overall research direction, planning, and integration among different research units. All authors contributed to the intellectual content and the writing of this work.

\section{Acknowledgment}
The work was supported by the National Research Foundation, Singapore under its Fellowship Award (NRF-NRFF13-2021-0010) and the Nanyang Technological University start-up grant (NTUSUG).  Work at Princeton University was supported by the Gordon and Betty Moore Foundation (GBMF4547 and GBMF9461; M.Z.H.).  J.Y.Y. and Y.P.F. are supported by the Ministry of Education, Singapore, under its MOE AcRF Tier 3 Award MOE2018-T3-1-002. The work at Northeastern University was supported by the US Department of Energy (DOE), Office of Science, Basic Energy Sciences Grant No. DE-SC0022216 and benefited from Northeastern University's Advanced Scientific Computation Center and the Discovery Cluster, and the National Energy Research Scientific Computing Center through DOE Grant No. DE-AC02-05CH11231.  

\suppinfo
Detailed structure information, structures, electronic properties, and Heisenberg models of antiferromagnetic kagome FeGe.


\clearpage

\end{document}